Check for updates

OPEN

# Interplay of the disorder and strain in gallium oxide

Alexander Azarov[1]✉, Vishnukanthan Venkatachalapathy[1,2], Platon Karaseov[3], Andrei Titov[3], Konstantin Karabeshkin[3], Andrei Struchkov[3] & Andrej Kuznetsov[1]✉

**Ion irradiation is a powerful tool to tune properties of semiconductors and, in particular, of gallium oxide ($Ga_2O_3$) which is a promising ultra-wide bandgap semiconductor exhibiting phase instability for high enough strain/disorder levels. In the present paper we observed an interesting interplay between the disorder and strain in monoclinic β-$Ga_2O_3$ single crystals by comparing atomic and cluster ion irradiations as well as atomic ions co-implants. The results obtained by a combination of the channeling technique, X-ray diffraction and theoretical calculations show that the disorder accumulation in β-$Ga_2O_3$ exhibits superlinear behavior as a function of the collision cascade density. Moreover, the level of strain in the implanted region can be engineered by changing the disorder conditions in the near surface layer. The results can be used for better understanding of the radiation effects in β-$Ga_2O_3$ and imply that disorder/strain interplay provides an additional degree of freedom to maintain desirable strain in $Ga_2O_3$, potentially applicable to modify the rate of the polymorphic transitions in this material.**

Ion implantation is a well-known technology to modify properties of materials. For example, it changed the paradigm in transistor manufacturing, providing accurate control of both spatial localization of dopants and their concentration[1]. Concurrently, the associated radiation disorder was considered as an artifact, rising a high demand for studies of the radiation disorder accumulation, with a practical focus to minimize the harm. More recently, a new concept of the defects functionalization became popular, e.g. along with the vision of the isolated intrinsic defects and defect complexes as building blocks for quantum devices[2–4]. Another defect functionalization option is to exploit the collective impact of the irradiation-induced disorder on the phase transitions occurring without changing chemical composition, i.e., polymorph transitions. In particular, such self-organized transitions have been recently observed in gallium oxide ($Ga_2O_3$)[5]. The phenomenon was interpreted in terms of the monoclinic (β-$Ga_2O_3$) to orthorhombic (κ-$Ga_2O_3$) or cubic (γ-$Ga_2O_3$) phase transition starting at a certain threshold of accumulated strain, in its turn controlled by the disorder level[5,6]. Notably, $Ga_2O_3$ is a very promising ultra-wide bandgap semiconductor for a range of applications and ion implantation is considered among the prime device fabrication tools[6–8]. Thus, on one hand the polymorphic transitions occurring under irradiation may be highly undesirable, since it may challenge the integrity of the device. On the other hand, the appearance of different polymorphs may turn to an advantage, if one gains control over the single-phase polymorph fabrication. In either case, the ultimate requirement is to understand and gain the control over the interplay between the disorder and strain in $Ga_2O_3$.

Notably, it is well established that the irradiation-induced disorder affects the strain level in the materials[9]. In semiconductors, the strain relaxation occurs most commonly via amorphization[10], while the polymorphic transitions are rare[5,11]. To the best of our knowledge, $Ga_2O_3$ is the only material where the irradiation induced polymorph transitions were demonstrated in a tractable way[5], in contrast to the observations in other materials demonstrating much less controllable polymorphism[11,12]. So far, the phenomenon in $Ga_2O_3$ has been studied as a function of the ion mass, dose, dose rate, and temperature[5,13,14]; altogether demonstrating a consistent trend for the strain accumulation induced by radiation defects.

However, there is yet no clear picture on how the disorder, and the corresponding strain accumulation, depend on the density of the collision cascades which is, in general, a rather complex function of the ion mass and energy[15]. Importantly, a direct comparison and unambiguous interpretation of the results obtained for irradiation regimes with ions having different atomic masses may be challenging. Previously, it was demonstrated that the cascade density effects can be efficiently studied by comparing the disorder induced by atomic and cluster ions, with the implantation parameters adjusted in such a way that the only difference between these irradiation regimes is the discrepancy between the atomic and cluster ions locations[16]; as schematically illustrated in Fig. 1. In the present paper, based on the formalism from Ref.[16], we compare all three types of implants, as illustrated in Fig. 1, to study the interplay between the disorder and strain in $\beta$-$Ga_2O_3$. As a result, we obtained a superlinear dependence of the disorder as a function of the collision cascades density and unveil the crucial role of the near-surface disorder for the strain accumulation. Such interplay between the disorder and strain provides an additional degree of freedom to maintain desirable strain in gallium oxide, potentially applicable to modify the rate of the polymorphic transitions in this material.

[1]Department of Physics, Centre for Materials Science and Nanotechnology, University of Oslo, Blindern, PO Box 1048, 0316 Oslo, Norway. [2]Department of Materials Science, National Research Nuclear University, "MEPhI", 31 Kashirskoe Hwy, 115409 Moscow, Russian Federation. [3]Peter the Great St.-Petersburg Polytechnic University, St.-Petersburg, Russia. ✉email: alexander.azarov@smn.uio.no; andrej.kuznetsov@fys.uio.no

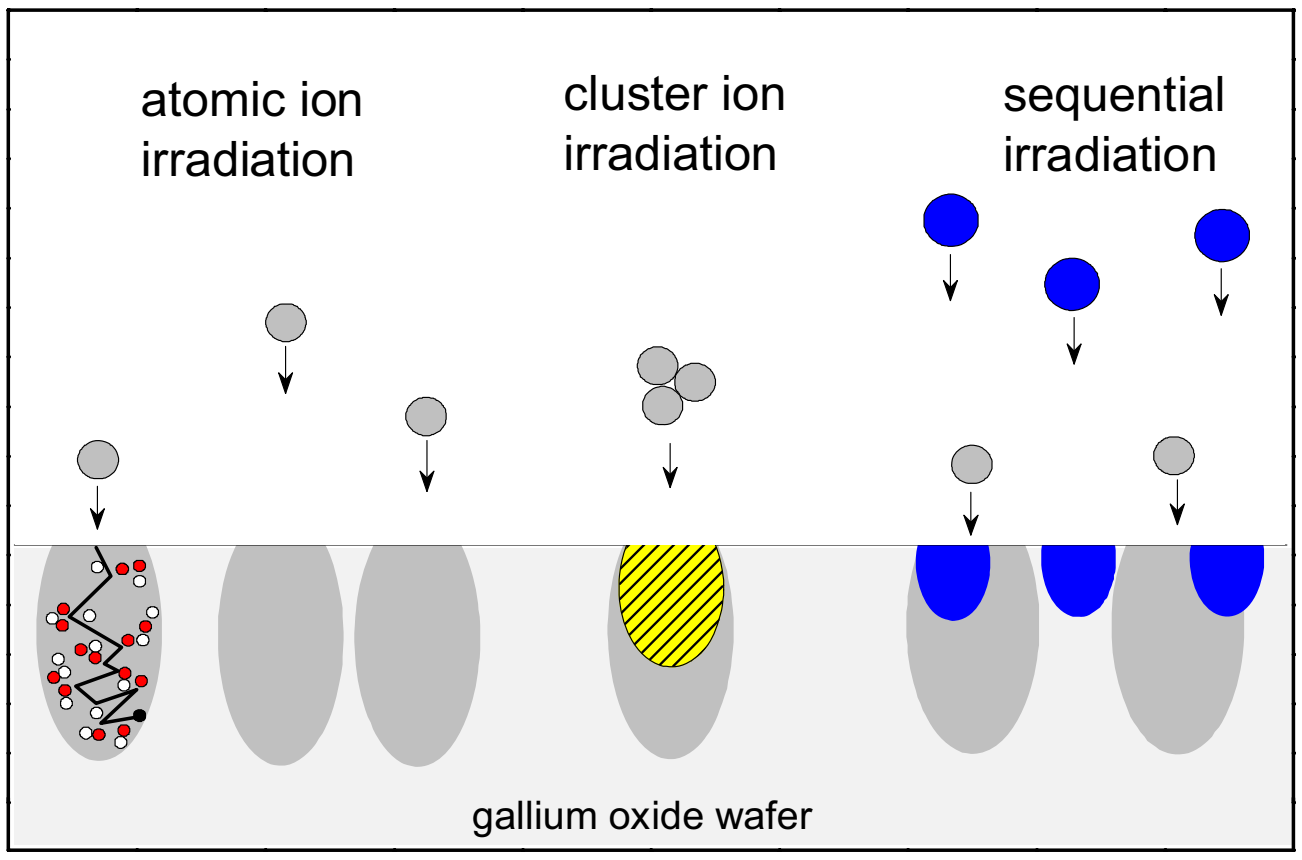

**Figure 1.** Schematics illustrating the difference between atomic, cluster and sequential (co-implants) ion irradiations. The yellow hatched area corresponds to overlapping of the individual collision cascades (with the light grey areas, containing defects (open and solid small circles) and an ion trajectory (black solid line) as shown for one of the cascades). The blue areas correspond to the collision cascades produced by low energy co-implants.

| | | Energy | | Dose | | Beam flux | |
|---|---|---|---|---|---|---|---|
| Ion | $R_{pd}$ (nm) | keV | keV/amu | ions/cm$^2$ | DPA | ions/(cm$^2$s) | DPA/s |
| $^{19}F^+$ | 46 | 60 | 3.2 | $1.9\times10^{14}$ | 0.15 | $3.1\times10^{12}$ | $2.5\times10^{-3}$ |
| $^{31}P^+$ | | 100 | | $1\times10^{14}$ | | $1.7\times10^{12}$ | |
| $^{69}PF_2^+$ | | 220 | | $5\times10^{13}$ | | $8.1\times10^{11}$ | |
| $^{107}PF_4^+$ | | 340 | | $3.3\times10^{13}$ | | $5.4\times10^{11}$ | |
| $^{58}Ni^+$ | 10 | 36 | 0.6 | $5\times10^{13}$ | 0.15 | $2\times10^{12}$ | $6\times10^{-3}$ |

**Table 1.** Implantation parameters used in the present study. The $R_{pd}$, DPA and DPA/s values were calculated using the SRIM code[18] simulations with the displacement energies of 25 and 28 eV for Ga and O atoms, respectively. The DPA values are obtained from the maximum of the vacancy profiles for a given dose normalized to the atomic density of $\beta$-$Ga_2O_3$ ($9.45\times10^{22}$ at/cm$^3$).

## Methods

In the present study (010) $\beta$-$Ga_2O_3$ single crystals were implanted with atomic (P and F) and cluster ($PF_2$ and $PF_4$) ions. The ion dose in displacements per atom (DPA), the defect generation rate in DPA/s, and the ion energy per atomic mass unit (amu) were kept the same for these implants. In addition, the low energy (36 keV) Ni ion implants were performed into the virgin sample as well as into the samples pre-implanted with F and P ions. Table 1 summarizes the energy, dose, beam flux as well as the depth corresponding to the nuclear energy loss maximum ($R_{pd}$) for all ions used. All implants were performed at room temperature, maintaining 7° off-angle orientation from normal direction to minimize channeling. The samples were analyzed by a combination of Rutherford backscattering spectrometry in channeling mode (RBS/C) and X-ray diffraction (XRD). RBS/C measurements was performed using 1.6 MeV $He^+$ ions incident along [010] direction and backscattered into a detector placed at 100° relative the incident beam direction. All RBS/C spectra were analyzed using one of the conventional algorithms[17] for extracting the effective number of scattering centres (referred to below as a 'relative disorder'). Thus, the relative disorder varies from 0 to 1 corresponding to the unimplanted and fully disordered states, respectively. XRD measurements were performed using the Bruker AXS D8 Discover diffractometer with high-resolution Cu $K_{\alpha1}$ radiation selected by a triple-bounce Ge (022) asymmetric monochromator exhibiting the instrumental broadness of 0.008°, as calibrated by Si (111) single crystal. The 2θ measurements were performed in the locked-coupled mode with position sensitive detector (PSD-LYNXEYE™) operating in the 1D mode.

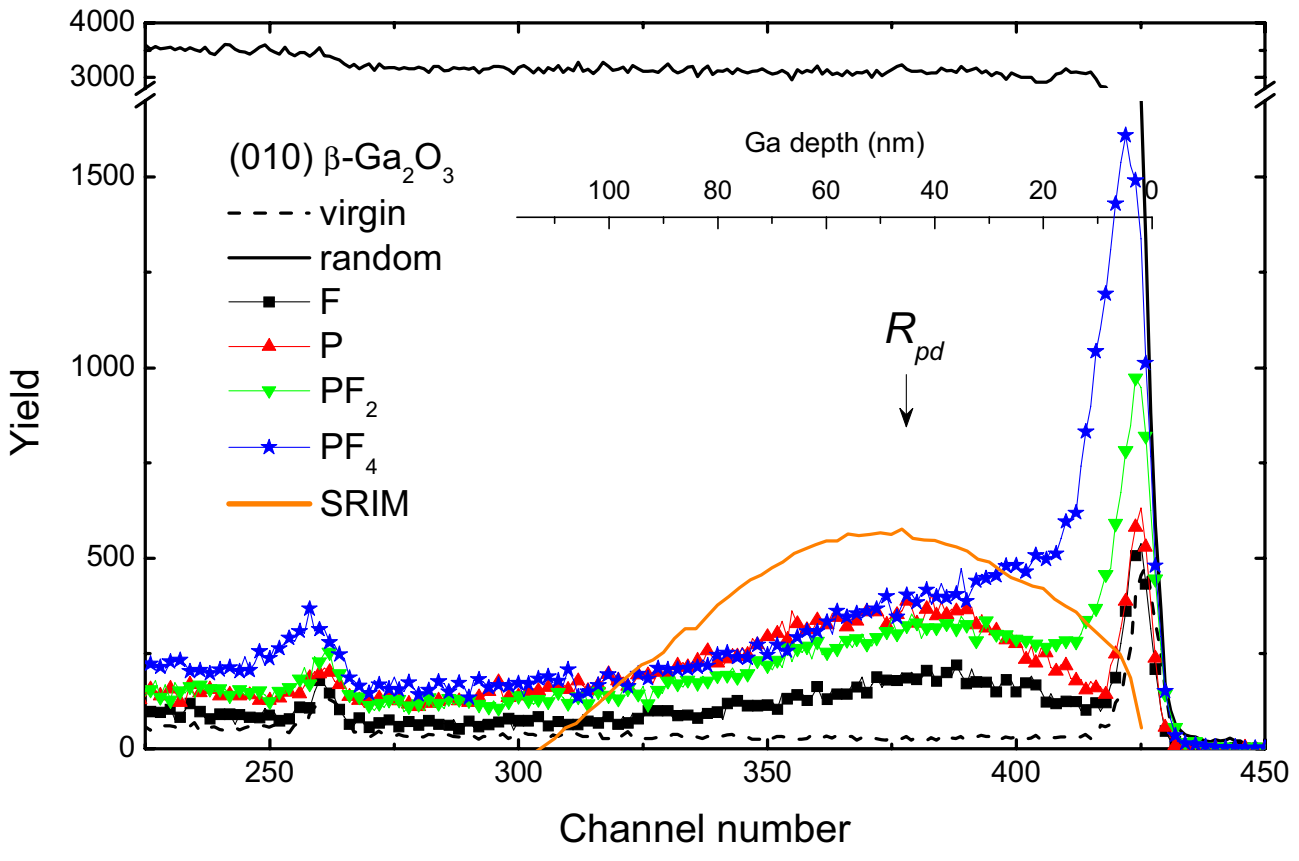

**Figure 2.** RBS/C spectra of (010) β-$Ga_2O_3$ implanted with atomic and cluster ions as indicated in the legend. For all implants the implantation conditions were chosen so that the ion doses in DPA and defect generation rates in DPA/s were kept identical based on SRIM simulations. The random and virgin (unimplanted) spectra are shown for comparison. The nuclear energy loss profile of P ions predicted with the SRIM simulation is also shown by the solid line in correlation with the Ga depth scale.

## Results and discussion

The role of the collision cascade density on the disorder formation in β-$Ga_2O_3$ is illustrated by Fig. 2 showing the RBS/C spectra of the samples implanted with atomic (F and P) and cluster ($PF_2$ and $PF_4$) ions. It should be noted, the sensitivity of the RBS technique is much better to heavy elements as compared to light atoms. Thus, in the rest of the paper we analyze the disorder in the Ga sublattice only; despite that the O signal is resolvable around 260 channel number in Fig. 2, corresponding to the surface position of O atoms. Thus, considering the Ga sublattice disorder, all implants in Fig. 2 result in two prominent features: one distinct peak at the surface and another broad peak centered at around $R_{pd}$; in direct comparison with the SRIM simulations data as indicated by the solid line in Fig. 2. Importantly, for the same DPA, P ions produce more defects in the bulk as compared to that of F which is a direct consequence of the collision cascade density effect ruled by the mass difference between these ions. In its turn, the cluster ion irradiation leads to a dramatic increase of the defects at the surface and in the region between the bulk and the surface defect peaks. It should be noted that even for $PF_4$ ions the height of the surface peak is well below the amorphous level, which is equivalent to the height of the random (fully disordered) spectrum. This surface disorder enhancement is attributed to the defect interaction from the overlapping collision cascades produced by the atoms comprising cluster ions as also schematically illustrated in Fig. 1. The cascade overlapping is obviously maximal near the surface, affecting the RBS/C data.

In order to better understand the mechanisms of the surface peak buildup, we use the SRIM code[18] simulations in combination with the previously developed methodology which takes into account the sub-cascades formation[19,20]. The results of the data processing are plotted in Fig. 3 (see "Supplementary Materials" for detailed description of the procedure). Notably, the inset in Fig. 3 shows the depth profiles of the effective cascade density as a function of the depth for all ions studied. It is seen that cluster ions produce much denser collision cascades for the depth ≤ 15 nm consistently with the data in Fig. 2. Thus, our results indicate a crucial role of the collision cascade density for the defect formation in β-$Ga_2O_3$. Indeed, Fig. 3 shows that the relative disorder in the surface peak exhibits superlinear behavior with increasing effective collision cascade density, with a very rapid disorder increase for the effective cascade density in excess of ~ 0.8 atomic %. Previously, it was demonstrated that surface amorphization in ion implanted GaN exhibits threshold-like behavior that was explained in the framework of the energy spike models[20]. However, in contrast to GaN, no amorphization was observed in ion implanted β-$Ga_2O_3$[5,7] and the disorder enhancement in β-$Ga_2O_3$ for cluster ions can be attributed to a nonlinear defect interaction as supported by a strong dose-rate effect in this material[13].

Notably, instead of amorphization, high dose irradiations result into a polymorphic transition in β-$Ga_2O_3$[5,14]. In this context, as it interpreted in[5], the strain accumulation in the implanted region is of paramount importance to control the polymorphic transitions. Thus, accounting for the data in Figs. 2 and 3, we investigated the interplay between the disorder and strain in our systematic set of samples as introduced by Fig. 1. The XRD data for the samples studied in Fig. 2 are plotted in Fig. 4 showing 2θ scans around the main (020) reflection (centered at 60.9°) of β-$Ga_2O_3$ (the full range XRD 2θ spectra are shown in Fig. S3 of "Supplementary Materials"). It is seen from Fig. 4 that for F implants, a high-angle shoulder appears on the side of the (020) reflection and this shoulder becomes more pronounced for P implants, resulting in a peak at ~ 61.04°. In context of the literature interpretations of the XRD data in irradiated materials, such shoulders/peaks may be attributed to the accumulation of compressive strain[21]. Previously, we demonstrated that this strain accumulation is scaling with the disorder level in the bulk damage peak[13]. However, surprisingly enough, the strain value does not scale up with the enhanced disorder in samples implanted with the cluster ions. Indeed, as clearly seen from Fig. 4 the intensity of the 61.04° peak is damped upon the $PF_2$ and $PF_4$ implants as compared to that for P implants. On the other hand the cluster implants lead to the formation of a broad tail at the right-hand-side of the (020) peak, see Fig. 4. This tail is most pronounced for $PF_4$ implants. Taking into account its strong asymmetry, it can be attributed to an enhanced concentration of extended defects[22]. Thus, lowering of the 61.04° peak for the cluster implants in contrast the P implants can be interpreted as shrinkage of the stressed layer due to strain relaxation stimulated by the extended defects in the near surface region.

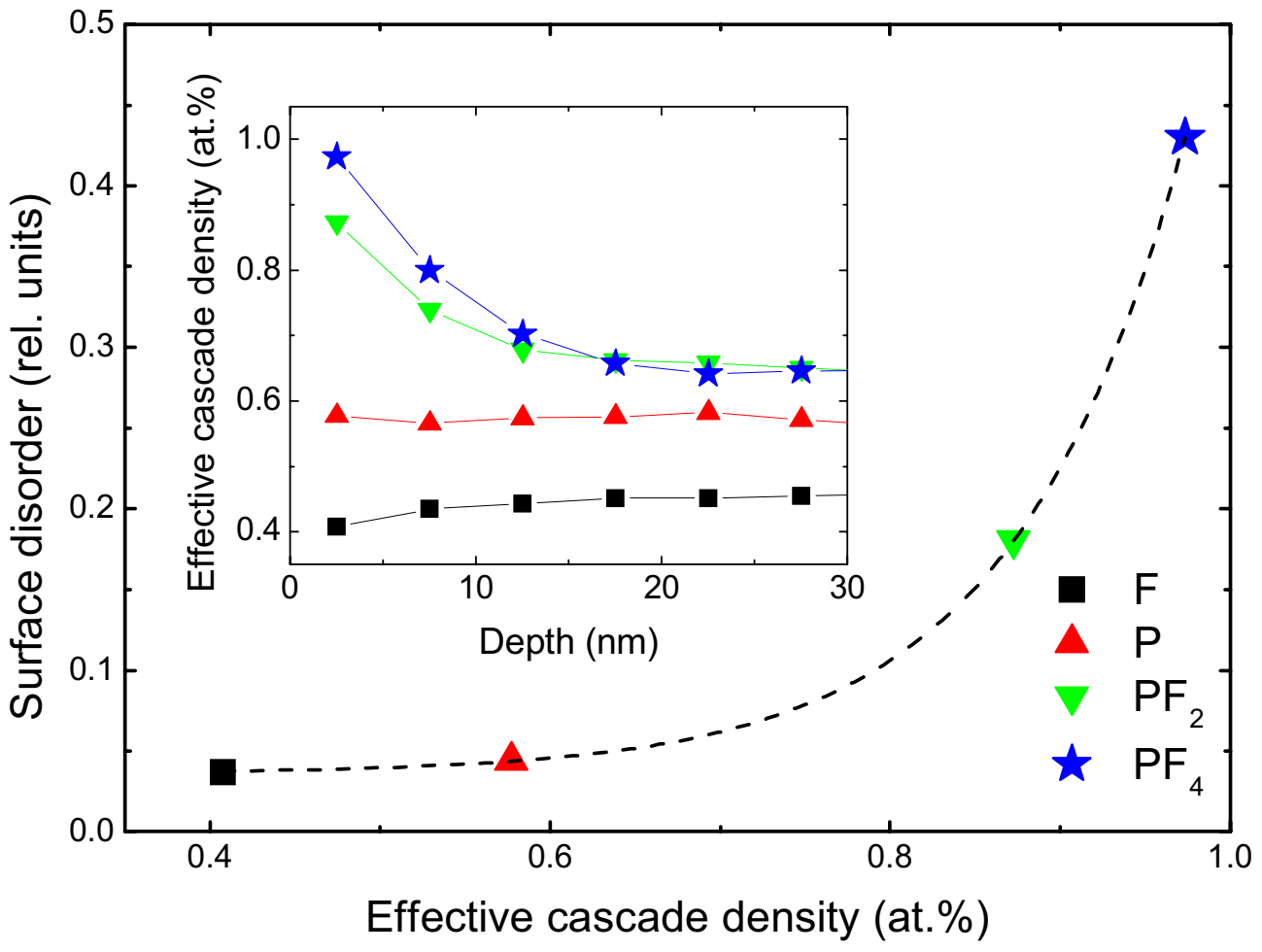

**Figure 3.** Amplitudes of the surface disorder peak in the ion implanted β-$Ga_2O_3$ (deduced from the spectra shown in Fig. 2) as a function of the effective cascade density for the atomic and cluster ion implants as indicated in the legend. The dashed line represents a superliner behavior and is to guide the reader's eye. The inset shows the evolution of the effective cascade densities as a function of depth.

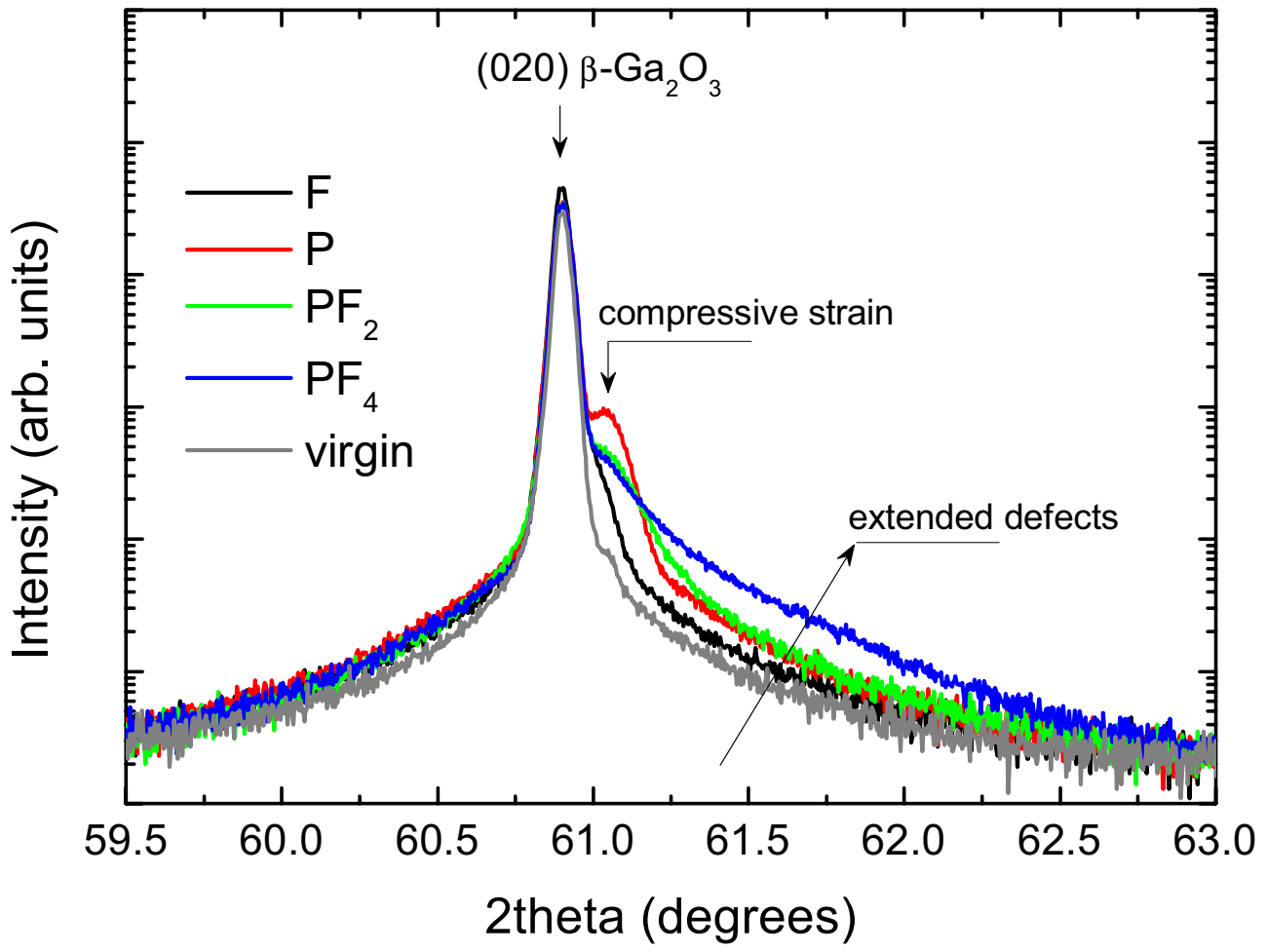

**Figure 4.** XRD 2theta scans across (020) reflection of the (010) β-$Ga_2O_3$ samples studies in Figs. 2 and 3.

The obtained results give an opportunity to manipulate the strain accumulation in the crystal bulk by the defect engineering in the near the surface region. In order to verify this hypothesis we performed additional low energy co-implantation with Ni ions into the samples having pre-existing disorder produced by F and P ions. The energy and the dose of Ni ions were chosen to produce similar disorder in the near surface region as in samples implanted with cluster ions (see the schematics in Fig. 1 and the implant parameters in Table 1). The impact of such co-implants on the interplay between the disorder and strain is illustrated by Fig. 5 including the disorder depth profiles, as deduced from the RBS/C spectra in panel (a) and the corresponding XRD data presented in panels (b) and (c). As expected, the low energy Ni implants generate disorder near the surface region, see Fig. 5a. Importantly, the disorder in the co-implanted samples is not a simple superposition of the individual disorder profiles. In particular, the amplitude of the surface disorder in the co-implanted samples is significantly higher as compared to the sum of the components, see Fig. 5a, consistently with nonlinear disorder accumulation for relatively high disorder levels[7,23]. Most importantly for the interplay with strain, we observe the decrease in the bulk disorder level in the co-implanted samples, systematically both for P + Ni and F + Ni co-implants. There are of course rational explanations for that, e.g. in terms the radiation-stimulated defect annealing due to migration of the defects from the near surface region and their interaction with the pre-existing disorder deeper in the bulk. It should be noted, that migration of the defects can be affected by the accumulated strain[24].

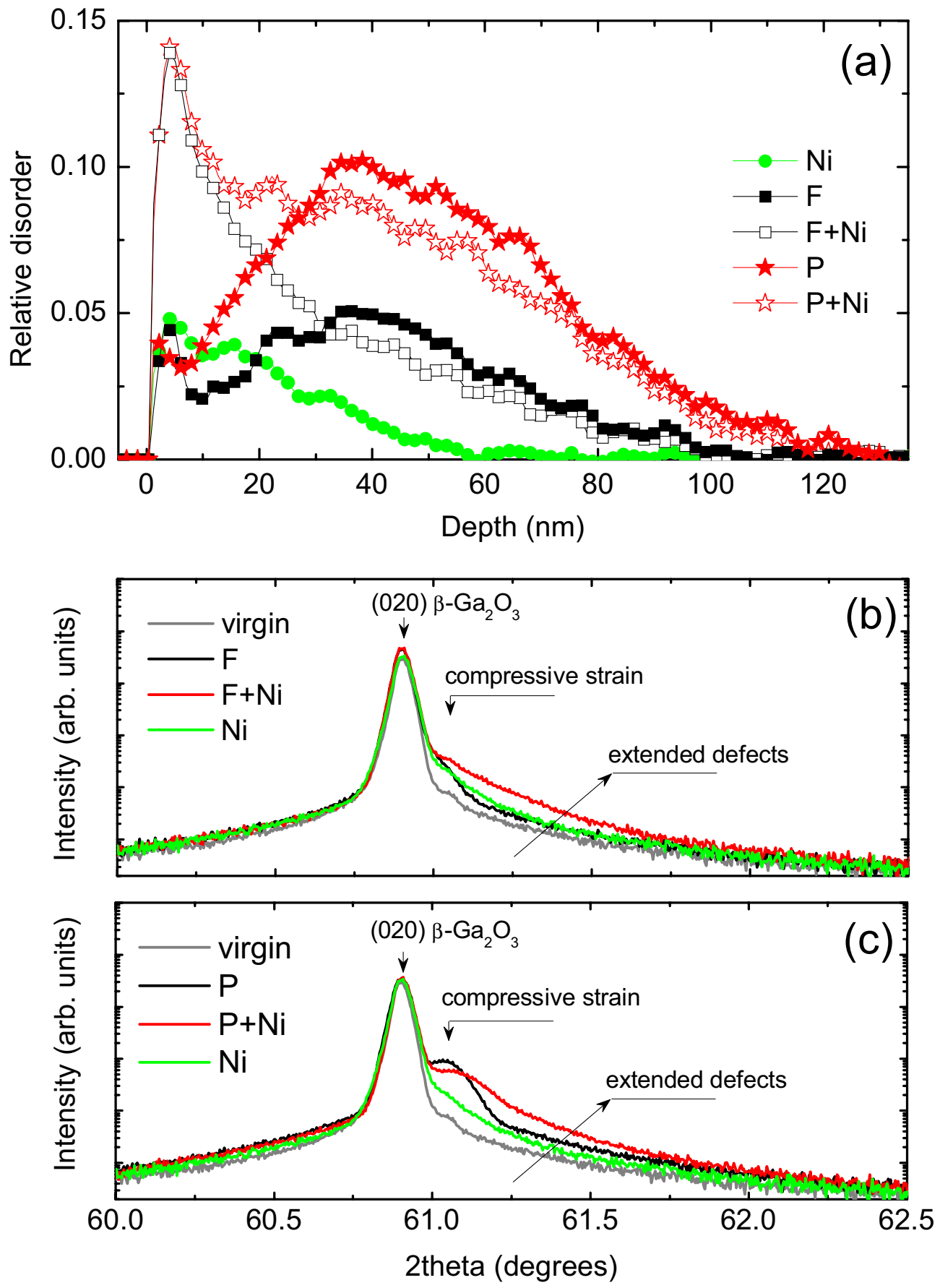

**Figure 5.** (**a**) Depth profiles of the relative for the co-implanted and single ion implanted (010) β-$Ga_2O_3$ samples; the corresponding XRD 2theta scans across (020) reflection are shown in the (**b**) and (**c**) for the sets of F and P ion implants, respectively.

The disorder changes in the co-implanted samples can be readily correlated with the strain evolution and, to some extent, with the corresponding features for the cluster ion implants. Indeed, one common feature is the extension of the tail on the high-angle side of the (020) reflection peak upon the co-implants, compare Figs. 4 and 5b,c. On the other hand, the co-implants do not release strain; e.g. for the P + Ni sample the 61.04° peak shifts to the higher angles indicating some increase of the compressive strain, as compared to the P implants alone. However, its intensity decreases that can be potentially attributed to a partial strain relaxation. Thus, even though the exact mechanisms affecting the difference in the disorder accumulation and strain evolution in β-$Ga_2O_3$ for implants with atomic ions having different mass, cluster ions, and co-implants may vary, there a consistent trend as shown in Figs. 2, 3, 4 and 5.

## Conclusions

In conclusion, we demonstrate that the collision cascade density has a strong impact on the ion-beam-induced defect/strain accumulation in gallium oxide. Specifically, we show that the disorder accumulation exhibits super-linear behavior with the collision cascade density. Moreover, the level of strain correlated with the amplitude of the bulk disorder peak can be engineered by changing the disorder conditions in the near surface layer. Such interplay between the disorder and strain was observed by comparing both atomic/cluster ion irradiations as well as atomic ion co-implants. Thus, the manipulation with the disorder depth profiles is an additional degree of freedom to maintain the desirable strain in gallium oxide, making it potentially applicable to modify the rate of the polymorphic transitions in this material.

## Data availability

All data generated or analysed during this study are included in this published article and its Supplementary Information file.

## Acknowledgements

The Research Council of Norway is acknowledged for the support to the Norwegian Micro- and Nano-Fabrication Facility, NorFab, project number 295864. M-ERA.NET GOFIB project (RCN project number 337627) is acknowledged for financial support. Work in St Petersburg was supported by the Russian Science Foundation (Project No. 22-19-00166). The INTPART Program at the Research Council of Norway, Projects Nos. 261574 and 322382, enabled the international collaboration.

## Author contributions

A.A., A.T. and A.K. designed research concept. A.A., V.V, K.K and A.S. carried out experiments. P.K. performed cascade density calculations. A.A. and A.K. wrote the main manuscript text and all the authors reviewed the manuscript.

## Competing interests

The authors declare no competing interests.

## Additional information

**Supplementary Information** The online version contains supplementary material available at https://doi.org/10.1038/s41598-022-19191-8.

**Correspondence** and requests for materials should be addressed to A.A. or A.K.

**Reprints and permissions information** is available at www.nature.com/reprints.

**Publisher's note** Springer Nature remains neutral with regard to jurisdictional claims in published maps and institutional affiliations.